\documentclass[a4paper,10pt]{article}

\title{Computing Functional and Relational
Box Consistency by Structured Propagation in
Atomic Constraint Systems
\thanks{
Presented at the Sixth Annual Workshop of the ERCIM Working Group
on Constraints, June 18-20, 2001, Charles University, Prague.
This document is
report DCS-266-IR, Department of Computer Science,
University of Victoria, Victoria, BC, Canada.}
}
\author{\mbox{M.H. van Emden}\\
        \mbox{Constraints Group, CWI
              and
              Aerospace Faculty, TU Delft
             }\\
        \mbox{On leave from Computer Science Dept,
              University of Victoria}
       }
\date{}

\newcommand{\C}{\mathcal{C}}
\newcommand{\E}{\mathcal{E}}

\newcommand{\R}{\mathcal{R}}

\newcommand{\T}{\mathcal{T}}
\newcommand{\ZA}{\hbox{{\scshape zero1}}}
\newcommand{\ZB}{\hbox{{\scshape zero2}}}

\begin{document}

\newtheorem{definition}{Definition}{}
\newtheorem{remark}{Remark}{}
\newtheorem{example}{Example}{}
\newtheorem{lemma}{Lemma}{}
\newtheorem{proposition}{Proposition}{}

\maketitle


\begin{abstract}
Box consistency has been observed to yield exponentially better
performance than chaotic constraint propagation in the interval
constraint system obtained by decomposing the original expression into
primitive constraints. The claim was made that the improvement is due
to avoiding decomposition. In this paper we argue that the improvement
is due to replacing chaotic iteration by a more structured
alternative.

To this end we distinguish the existing notion of box consistency from
\emph{relational} box consistency.  We show that from a computational
point of view it is important to maintain the functional structure in
constraint systems that are associated with a system of equations.  So
far, it has only been considered computationally important that
constraint propagation be fair.  With the additional structure of
functional constraint systems, one can define and implement
computationally effective, structured, truncated constraint
propagations. The existing algorithm for box consistency is one such.
Our results suggest that there are others worth investigating.

\end{abstract}

\section{Introduction}
\vspace{-1mm}

Systems of nonlinear equations where the unknowns are reals arise in
specialized applications such as the study of chemical equilibrium and
in robot kinematics.  A general class of applications of systems of
nonlinear equations arises when optimizing a function of $n$ variables
with multiple local minima.  A common optimization method is to set the
$n$ partial derivatives to zero, and solve the resulting set of $n$
equations, which are in general nonlinear.
Thus, nonlinear equations occur widely in mathematical modeling.

Until recently, only Newton's method was available for solving such a
system.  This method is good at refining sufficiently good estimates
of a solution.  Trying to use it otherwise is a hit-and-miss affair.
The situation was greatly improved with the advent of interval
arithmetic
\cite{moore66,nmr90}.

A remarkable subsequent development was BNR Prolog \cite{BNR88} (later
referred to as ``CLP(Intervals)''), which introduced what came to be
called \emph{interval constraints}.  This method can be regarded as an
adaptation of the CHIP system \cite{dvhsagb88}.  This constraint
processing system associates each variable with a finite set of
possible values. Instead, in BNR Prolog, the set of possible values is
an interval of reals.  BNR Prolog adopted from CHIP an instance of a
constraint propagation algorithm that turned out to be an instance of
Apt's Generic Chaotic Iteration algorithm (``GCI'' in the sequel)
\cite{aptEssence}.

To solve nonlinear systems with CLP(Intervals), one has the advantage
of not needing derivatives.  The equations are decomposed into
primitive constraints.  The resulting constraint system is then
subjected to a constraint propagation algorithm (the
\emph{pruning step}).  The resulting box is split (the \emph{branching
step}), whereupon the same is done recursively to the results of the
split.  By ensuring that pruning preserves completeness (that is, does
not remove any solutions), one ensures that the solving
algorithm generates a sequence of boxes that contain all solutions.

Benhamou, McAllester, and Van Hentenryck \cite{bmcv94} showed that
this simplicity comes at a cost: on the Broyden Banded Function, a
widely used benchmark, exponentially increasing computation time is
needed when one only uses constraint propagation for pruning.  This is
not surprising because of the presence of branching.

By contrast \cite{bmcv94} describes {\tt Newton}, an algorithm that
does not require branching in this benchmark and only exhibits
linearly increasing computation time. This remarkable improvement was
based on the novel notions of \emph{box consistency} and of the
\emph{pseudo-zeros} used to characterize maximally box-consistent sets.

{\tt Newton} avoided branching on this particular benchmark because
box-consistency achieved stronger pruning than the constraint
propagation used in CLP(Intervals). However, box consistency was a
step back in the sense of only using functional interval arithmetic,
which contracts only the interval for the function value, rather than
using the propagation of CLP(Intervals), which has the potential of
contracting all intervals involved in a relation. Logically, the next
improvement was to be one that combined the advantages of box
consistency with those of the relational interval arithmetic used in
CLP(Intervals). This step was taken by Benhamou, Goualard, Granville,
and Puget with their HC4 algorithm \cite{bggp99}.

In this paper, we explore other ideas for improving the use of
relational interval arithmetic. We describe the use of \emph{probing}
an interval constraint system to improve the bounds obtained by a
single constraint propagation. We show that this leads to a relational
form of box consistency, which is stronger than the original notion of
box consistency, which we call \emph{functional} box consistency. We
show that functional and relational box consistency are but two
extremes of a spectrum defined by ways of structuring and truncating
the iteration in constraint propagation. One of these ways can be
regarded as the simulation of the evaluation of an expression in
interval arithmetic.

\paragraph{Disclaimer}
Many basic definitions and results need to be covered here. In most
cases, no attempt at attribution will be made.  In the interest of
mutual compatibility, some definitions are modified. As a result
attribution might not be welcomed, yet no novelty is involved.
Possible novelties are simulation of interval arithmetic by constraint
propagation, relational box consistency and its computation by
probing, and the identification of alternatives to full constraint
propagation based on structured rather than chaotic iteration.

\section{Constraints and equations}
\label{constrEq}
\vspace{-1mm}

\subsection{Constraint systems}
\vspace{-1mm}
\begin{definition}
\label{constrSys}
A constraint system has the following attributes.\\
(1) A set $\{T_1,
\ldots, T_n\}$ of sets called \emph{types}. \\
(2) A set $\{x_1,
\ldots, x_n\}$ of variables where $x_i$ is of type ${T_i}$ for $i \in
\{1,\ldots,n\}$.\\
(3) A set $\{A_1, \ldots, A_m\}$ of constraints
where $A_i$ is an atomic formula of first-order predicate logic.
$\{x_1, \ldots, x_n\}$ is the set of all variables occurring in
$\{A_1, \ldots, A_m\}$.\\ For $i=1,\ldots,m$, $d_i \subseteq
\{1,\ldots,n\}$ is such that $\{x_j \mid j \in d_i \}$ is the
\emph{set of variables occurring in} $A_i$.\\
(4) A \emph{state},
which is $D_1 \times \cdots \times D_n$, where, for $i \in
\{1,\ldots,n\}$, $D_i \subseteq T_i$.  We say that, in this state of
the constraint system, $D_i$ is the \emph{domain} of $x_i$.\\
(5) An
\emph{initial state}, which is a state.
\end{definition}

\begin{definition}
The \emph{relation  associated with} a function or operation
$f: \R^k \to \R$ is 
$\{\langle x_0, \ldots, x_k \rangle \mid x_0 = f(x_1, \ldots, x_k) \}$,
for $k = 0,1,\ldots$
\end{definition}

We will say that a function or operation is \emph{admissible} if the
contraction operator of the
associated relation is efficiently
computable.  This means roughly that it can be computed without
iteration.  Admissible operations include addition, subtraction,
multiplication, division, maximum, absolute value,
power for all integer exponents,
exp, log, and the trigonometric functions.

\begin{proposition} \label{atomicSim}
Let $r$ be the relation associated with an admissible $f : \R^k
\rightarrow \R$.  Let the intervals associated with
$x_0,x_1,\ldots,x_k$ be $[-\infty,+\infty],I_1,\ldots,I_k$,
respectively. Then the intervals resulting from applying the
contraction operator of $r$ are, respectively, the intervals
$f^\prime(I_1,\ldots,I_k),I_1,\ldots,I_k$, where $f^\prime$ is the
canonical set extension of $f$.
\end{proposition}

\begin{definition}
An \emph{equation} is $E=0$, where $E$ is an expression of type real
containing only real variables. An \emph{equation system} consists of a
set $X$ of real variables and a set of equations containing no
variables other than those in $X$.
\end{definition}

In the conventional way, we consider expressions as trees. The leaf
nodes are constants or variables; the nonleaf nodes are the operation
symbols.

A distinguishing feature of CLP(Intervals) is that it decomposes
equations, or other composite expressions, into primitive
constraints. These primitive constraints are the relational versions
of the building blocks of expressions, which are admissible functions.

It is this decomposition, described in the definition below, that has
been identified in \cite{bmcv94} as the cause for the observed
slowness of CLP(Intervals).

\begin{definition}
The \emph{constraint system} $\C$ \emph{associated} with an equation system
$\E$ depends on a one-one correspondence between the non-leaf
nodes of the trees in $\E$ and a set of variables that is disjoint
from the variables in $\E$, which is defined as follows.

Each of the variables in $\E$ also occurs in $\C$, where it is called a
``primary variable''. To each non-leaf node of an expression in $\E$
there corresponds a variable in $\C$ that does not occur
in $\E$ and is called ``auxiliary variable''.

The constraints in $\C$ are determined as follows.  For every non-leaf
node $n$ (which is an operation $f$), with children $n_1,\ldots,n_k$,
of a tree in $\E$, there is an atomic formula in $\C$ with variables
$x_0,x_1,\ldots,x_k$.  The predicate in the atomic formula denotes the
relation associated with $f$.  This formula is a \emph{functional
atom} in $\C$.  The variables $x_1,\ldots,x_k$ are the \emph{input
variables} of the atom; $x_0$ is its \emph{output variable}.

In addition there is in $\C$, for every root of a tree in $\E$ with
corresponding variable $v$, an atomic formula $v=0$.
This is a \emph{relational atom} in $\C$.
\end{definition}

\begin{definition}[Floating-point numbers, intervals]
A floating-point number is any element of $F \cup \{-\infty,+\infty\}$,
where $F$ is a finite set of reals that includes $0$.
If $x$ is a finite floating-point number,
then $x^-$ ($x^+$) is the greatest (smallest)
floating-point number smaller (greater) than $x$.
In addition,
$-\infty^- = -\infty$,
$-\infty^+ = -M$,
$+\infty^- = M$,
and
$+\infty^+ = +\infty$,
where $M$ is the greatest finite floating-point number.

A \emph{floating-point interval}
is a closed connected set of reals, where the bounds, in so far as they
exist, are floating-point numbers.  When we write ``interval'' without
qualification, we mean floating-point interval.

An interval that does not properly contain an
interval is called \emph{canonical}.

A \emph{box} is a cartesian product of floating-point intervals.
\end{definition}

Thus canonical intervals are non-empty sets of reals.  They may have
positive width and they may have zero width.
Examples are $[a^-,a]$, $[a,a^+]$, and $[a,a]$, where $a$ is a finite
float-point number.
For any real, there is a unique smallest canonical
floating-point interval containing it.

In this paper we consider \emph{interval constraint systems}, which
are constraint systems where the types are all equal to $\R$ and where
the domains are intervals.

\section{Propagation}
\label{propg}
\vspace{-1mm}

Apt's Generic Chaotic Iteration algorithm (GCI) \cite{aptEssence} is of
an astonishing simplicity and wide applicability.
The elegance of CLP(Intervals), noted in \cite{bmcv94},
is due in part to the fact that its constraint propagation is an
instance of GCI.

GCI maintains a pool of operators that still need to be applied.  The
attraction of GCI is that it does not specify any order among these
applications.  In this section we consider orders of application that
are computationally effective for constraint systems that are
associated with equation systems.  We formalize application order by
means of ``traces'', as defined below.

\begin{definition}
A \emph{trace} for a given constraint system $\C$ with attributes
as in Definition~\ref{constrSys} has the following
components:

(1) An \emph{index sequence} $t$, which is an infinite sequence with
elements in $\{1,\ldots,m\}$.

(2) A \emph{sequence-of-atoms} of which the $i$-th element is the
atom $A_{t_i}$ in $\C$, for $i=0,1,\ldots$

(3) A \emph{sequence-of-constraints} of which the $i$-th element is the
relation defined by the atom $A_{t_i}$ in $\C$, for $i=0,1,\ldots$

(4) A \emph{sequence-of-contraction operators} of which the $i$-th element
is the contraction operator $\tau_{t_i}$
defined by the atom $A_{t_i}$ in $\C$,
for $i=0,1,\ldots$

(5) A \emph{sequence-of-boxes} $U$ of which the $i$-th element is
the initial box of $\C$
if $i=0$ and is $\tau_{t_{i-1}}(U_{i-1})$ if $i>0$.
\end{definition}

As we are primarily interested in the sequence of boxes, we think of
the sequence of contraction operators as ``activations'' of the
corresponding constraints. We think of the elements of $t$ as
``selecting'' a constraint to be activated.

The following proposition is based on the fact that
the contraction operators are monotone nonincreasing and idempotent
and that there are finitely many domains.

\begin{proposition} 
See \cite{vnmd97,aptEssence}.                  \\
For any interval constraint system with
box $B$ as initial state we have:\\
(1)
The sequence of boxes has a limit for every trace.\\
(2)
The greatest lower bound of these limits is also a limit of a trace.\\
(3)
All traces in which the index sequence is fair have
the same limit. This limit equals the greatest common fixpoint of
$\tau_1,\ldots,\tau_m$ that is less than $B$.\\
(4)
This fixpoint is uniquely  determined by the constraint system.
It is computed by a suitable instance of GCI.
\end{proposition}

\begin{definition}
A constraint system is \emph{failed} (\emph{non-failed}) if its
fixpoint is empty (non-empty).
\end{definition}

A failed constraint system has no solutions.
A non-failed constraint system may, but need not, have solutions.

\begin{proposition}
The fixpoint of a constraint system contains all its solutions.
\end{proposition}

\begin{definition}
A segment of a trace is \emph{functional} if

(1) The sequence-of-atoms only contains functional atoms.

(2) For every atom, any input variable that is an auxiliary variable
has occurred as output variable earlier in the segment.

(3) No atom occurs more than once.

A segment of a trace is \emph{inverse functional} if every occurrence
of a variable as an output variable has been preceded by an
occurrence of that variable as input variable or as variable in an
activation of an equality constraint.
\end{definition}

The following proposition shows that a trace of a constraint system
can simulate the evaluation of an expression.

\begin{proposition}
\label{simul}
Let $\C$ be an interval constraint system
with attributes as in Definition~\ref{constrSys} that is
associated with an equation system
$\E$ containing an expression $E$.
Let box $B$ be the initial state of $\C$ such that all its projections
corresponding to auxiliary variables are $[-\infty,+\infty]$.
Let $d \subseteq \{1,\ldots,n\}$ be such that
$\{x_j \mid j \in d\}$ is the set of variables in $E$.
Let $x_i$ be the variable in
$\C$ that corresponds to the root of $E$.
Let $\T$ be
a functional initial segment of a trace.

The $i$-th projection of the last of the sequence-of-boxes of $\T$
is the same interval as the one obtained
when $E$ is evaluated in interval arithmetic
with $\pi_j(B)$ as the interval substituted for $x_j$ in $E$, for all
$j \in d$. 
\end{proposition}


\paragraph{Consequences of Proposition~\ref{simul}}
GCI leaves open the possibility of activating an operator that has no
effect; that is, when this activation does not result in any domain
reduction.
In any trace containing an operator activation without effect, the number
of steps to convergence can be decreased by removing it.

One possible heuristic to avoid vacuous activation is to require a
trace to be a \emph{two-phase iteration}\/: to consist of a repetition
of cycles consisting of the following two items: (1) a functional
segment and (2) an inverse-functional segment.

It may be conjectured that such a trace is optimal in the sense of
having a shortest pre-convergence initial segment.  It is to be expected
that such a trace is subject to severely diminishing returns in the
sense of activations resulting in domain reduction.

The first functional segment likely has the greatest effect. It
corresponds to the interval-arithmetical special case of constraint
propagation. While one can probably construct examples to the
contrary, most of the remaining reduction is typically effected in the
first inverse-functional segment.  For some purposes, to be discussed
below, it is advisable to truncate constraint propagation before
convergence has occurred. Promising truncations are: after the first
functional segment or after the first cycle.

\section{Box consistency}
\label{boxc}
\vspace{-1mm}

We first describe, adapted to the current setting, the box consistency
notion of \cite{bmcv94}. To distinguish it from the version to be
described next, we call it ``functional box consistency''.

\subsection{Functional box consistency}
\label{funcBoxCon}
\vspace{-1mm}

\begin{definition}
{\bf (Coordinate-wise functional box consistency operator)}\\
Given an equation $E = 0$ with variables $x_1,\ldots,x_n$,
and $i \in \{1,\ldots,n\}$.
The $i$-th
\emph{coordinate-wise functional box consistency operator}
replaces in $I_1 \times \cdots \times I_n$ the interval $I_i$ by
its intersection with the smallest interval containing
$\{ y \in \R \mid 0 \in E^\prime \}$, where
$E^\prime$ is the result of evaluating $E$ in interval arithmetic with
$I_1,\ldots,I_n$ substituted for the variables
$x_1,\ldots,x_n$, except that
the smallest floating-point interval containing $y$
is substituted for $x_i$.
\end{definition}

The $i$-th coordinate-wise functional box consistency operator is
completeness-preserving in the sense that it removes no solution to
$E=0$.

\begin{definition}[Functional box consistency]
\label{FBC}
A box is \emph{functionally box consistent} with respect to an equation
if it is a common fixpoint of the $n$ coordinate-wise functional box
consistency operators associated with that equation.
\end{definition}

This is equivalent to the notion of box consistency introduced in
\cite{bmcv94}.  By itself, box consistency is not interesting: for
example, the empty box is box-consistent with respect to $\{x=0\}$,
which has plenty of zeros.  What is lacking so far is a suitable notion
of maximality.  This was done in \cite{bmcv94} by relating box
consistency with leftmost and rightmost pseudo-zeros.  It can also be
done by means of fixpoint theory.  Because each coordinate-wise
functional box consistency operator is a monotonic one defined on the
partially ordered set of boxes with set containment as partial order,
it has a greatest fixpoint.  By applying GCI one obtains the greatest
fixpoint common to all $n$ coordinate-wise functional box consistency
operators associated with a set of equations.

\begin{definition}[Functional box consistency operator]
\label{FBCO}
Given an equation $E = 0$ with variables $x_1,\ldots,x_n$.  The
\emph{functional box consistency operator} is the mapping from a box
$B$ to the greatest fixpoint contained in $B$ that is common to all $n$
coordinate-wise functional box consistency operators associated with
the equation.
\end{definition}


\begin{definition}[Functional pseudo-zero]
\label{LRPZ}
Given an expression $E$ with variables $x_1,\ldots,x_n$.  An
\emph{i-th functional pseudo-zero} of $E$ with respect to the
intervals $\{I_j \mid j \in \{1,\ldots,n\}\setminus\{i\}\}$ is a
canonical interval $I$ such that $0$ is contained in the interval
resulting from evaluating $E$ in interval arithmetic with $I$
substituted for $x_i$ and $I_j$ substituted for $x_j$, for all $j \in
\{1,\ldots,n\}\setminus\{i\}\}$. \\
An \emph{$i$-th functional pseudo-zero of an equation system} with
respect to the intervals $\{I_j \mid j \in
\{1,\ldots,n\}\setminus\{i\}\}$ is a canonical interval that is a
functional pseudo-zero of every one of its equations.

\end{definition}

\begin{proposition}
Given an  expression $E$ with variables $x_1,\ldots,x_n$.
If, for any $i \in \{1,\ldots,n\}$, no $i$-th functional pseudo-zero
exists with respect to
$\{I_j \mid j \in \{1,\ldots,n\}\setminus\{i\}\}$,
then $E=0$ has no solution with
$x_j \in I_j$, for all $j \in \{1,\ldots,n\}\setminus\{i\}\}$.
\end{proposition}

\begin{proposition}
The result of applying the functional box consistency operator of an
equation system with at least one solution to a box $B$ results in a
box that has as $i$-th projection the least
interval containing the $i$-th
leftmost and rightmost functional pseudo-zeros with respect to
$\pi_1(B),\ldots,\pi_n(B)$, for all $i \in \{1,\ldots,n\}$.
\end{proposition}

\subsection{Probing}
\vspace{-1mm}

\begin{definition}
Let $\C$ be a constraint system.  \emph{Probing} $\C$ \emph{with a
constraint} $A$ means determining whether the constraint system
$\C^\prime$, which is the result of adding $A$ to $\C$, is failed or
non-failed.
\end{definition}

Note that $\C$ does not change as the result of probing.

\begin{proposition} [Monotonicity of probing]
\label{monProb}
If a non-failed constraint system $\C$ yields failure as a result of
probing with $x \leq u_1$, then it also yields failure as a result of
probing with $x \leq u_2$, where $u_1$ and $u_2$ are floating-point
numbers such that $u_2 < u_1$.
\end{proposition}
A similar fact holds for probing with $x \geq l_1$ or
$x \geq l_2$, with $l_2 > l_1$. Also non-failure can be inferred
from non-failure by means of suitably selected probes.

When probing a constraint system $\C$ containing a variable $x$ with
$x \leq a$ yields failure, then it has been proved that no solution
has an $x$-component that is less than or equal to $a$. It does not
follow that probing $\C$ with $x \geq a$ yields non-failure.  It is
quite common for $\C$ to be non-failed, yet not to have any
solutions. This is sometimes discovered by probing $\C$ with both $x
\leq a$ and $x \geq a$ and finding failure in both cases.  While this
property of probing is often effective, such an $x$ and such an $a$
cannot always be found.

As seen above, probing has a logic of its own.
Here is an other example. It may be that  a non-failed
constraint system $\C$ containing variable $x$
yields non-failure on probing with $x \geq a$
and that it
yields non-failure on probing with $x \leq b$,
where $a < b$, and yet yields failure on probing with
$x \geq a \wedge x \leq b$.
Conversely, non-failure on probing with 
$x \geq a \wedge x \leq b$
implies non-failure on probing with $x \geq a$;
it also
implies non-failure on probing with $x \leq b$.

Probing suggests a relational version of the functional
pseudo-zero. As probing applies to any constraint system, not
just to those derived from equations, we call them
``pseudo-solutions''.

\subsection{Relational box consistency}
\vspace{-1mm}

Probing can be used to compute approximations to relational box
consistency, a criterion similar to functional box consistency, but
one that produces closer approximations to the set of
solutions. Another way in which relational box consistency is
interesting is that it applies to all interval constraint systems, not
only to those that are derived from equations.

We proceed in analogy with Section~\ref{funcBoxCon}. In analogy with
the pseudo-zeros of \cite{vhlmyd97} we have the following definition.

\begin{definition}[(Canonical) pseudo-solution]
Let $\C$ be a non-failed interval constraint system with initial box
$I_1\times \cdots \times I_n$ and let $i \in \{1,\ldots,n\}$.  Let
$\C^\prime$ be an interval constraint system with the same attributes as
$\C$, except that interval $I_i$ is changed to an interval $y$.  If
$\C^\prime$ is non-failed, then y is an \emph{i-th pseudo-solution of
$\C$ with respect to}
$\{I_j \mid j \in \{1,\ldots,n\}\setminus\{i\}\}$.
If $y$ is a canonical interval, then it is an \emph{i-th
canonical pseudo-solution of $\C$ with respect to} $\{I_j \mid
j \in \{1,\ldots,n\}\setminus\{i\}\}$.
\end{definition}

Functional pseudo-zeros have been defined, following \cite{bmcv94,
vhlmyd97} as being canonical intervals. Here it is useful that
pseudo-solutions are not necessarily canonical.

\begin{definition}[Coordinate-wise relational box consistency operator]
The \emph{$i$-th coordinate-wise relational box consistency operator}
of an interval constraint system $\C$ maps the initial box $I_1\times
\cdots \times I_n$ to one where the $i$-th projection ($i \in
\{1,\ldots,n\}$) has been replaced by the least floating-point
interval containing the union of the $i$-th canonical pseudo-solutions
of $\C$ with respect to
$\{I_j \mid j \in \{1,\ldots,n\}\setminus\{i\}\}$.
\end{definition}

We introduced Definition~\ref{FBC} because it figures so
prominently in the literature, following \cite{bmcv94}. One could
define a relational analog, but it is more useful to skip to the
analogs of Definition~\ref{FBCO} and Definition~\ref{LRPZ}.

\begin{definition}[Relational box consistency operator]
Given an interval constraint system $\C$. The \emph{relational box
consistency operator} of $\C$ maps its initial box $I_1 \times \cdots
\times I_n$ to the greatest fixpoint contained in it that is common to
all $n$ coordinate-wise relational box consistency operators
associated with $\C$.
\end{definition}

\begin{definition}[Leftmost (Rightmost) pseudo-solution]
Let $\C$ be an interval constraint system with variables
$x_1,\ldots,x_n$ and let $i \in \{1,\ldots,n\}$. An $i$-th
pseudo-solution $[a,b]$ of $\C$ is an \emph{$i$-th leftmost
pseudo-solution of} $\C$ if $\C$ has no solution $\langle
\xi_1,\ldots,\xi_n \rangle$ with $\xi_i$ in $\{ x \in \R \mid x \leq y
\hbox{ for all } y \in [a,b] \}$\footnote{ It is simpler to say:
``with $\xi_i \leq a$'', but this does not yield a useful definition
because $\xi_i$ is a real and $a$ is a floating-point number. As a
result, $a$ can be $-\infty$.  }.
\end{definition}

Pseudo-solutions become less interesting the wider they are. For
example, $[-\infty,+\infty]$ is a pseudo-solution, a leftmost, and a
rightmost pseudo-solution for any interval constraint system. On the
other hand, consider an $i$-th canonical pseudo-solution $[a,b]$ such
that no $i$-th canonical pseudo-solution is to the left of it. As
every real $\xi_i$ that is the $i$-th component of a solution is
contained in an $i$-th canonical pseudo-solution, $[a,b]$ is also an
$i$-th leftmost pseudo-solution. Hence the following proposition.

\begin{proposition}
Suppose the relational box consistency operator maps box $B$ to
$B^\prime$. Then $B^\prime$ is equal to the smallest box containing
the $i$-th leftmost and rightmost canonical pseudo-solutions, for $i
\in \{1,\ldots,n\}$.
\end{proposition}

The reason for having these different characterizations for the same
object is that the one in terms of pseudo-zeros is convenient for a
relational box-consistency algorithm; see
section~\ref{algSection}. The one in terms of fixpoints is convenient
for comparing relational box consistency with functional box
consistency.

\begin{proposition}
Let $E$ be an expression with $n$ variables and let $\C$ be the
interval constraint system associated with $\{E=0\}$. The result of
mapping a box $B$ with the relational box consistency operator of $\C$
is contained in the result of mapping $B$ with the functional box
consistency operator of $\{E=0\}$.
\end{proposition}

{\bf Proof.}  Both mapping results are characterized as common
greatest fixpoints. Hence it is sufficient to show that, with respect
to $\{I_j \mid j \in \{1,\ldots,n\}\setminus\{i\} \}$,
the result of
the $i$-th relational box consistency operator is contained in that of
the $i$-th functional box consistency operator.

This can be ascertained by considering a canonical interval $[a,b]$
for which $0 \not\in f_i([a,b])$. Here $f_i$ is a function from
intervals to intervals such that $f_i(I)$ is the result of evaluating
$E$ in interval arithmetic with $I_j$ substituted for $x_j$ for $j \in
\{1,\ldots,n\}$ except that $I$ is substituted for $x_i$.

Such an interval $[a,b]$ lies outside the
result of applying the
$i$-th functional box
consistency operator. One fair trace of $\C$ begins with a functional
segment followed by activating the equality constraint. It yields the
empty interval at that point, which ensures that the unique limit of
any fair trace is empty. Hence $[a,b]$ also lies outside the result
of the $i$-th relational box consistency operator.

\section{Computing Box Consistency}
\label{algSection}
\vspace{-1mm}

As only maximal box-consistent boxes are of interest, we compute leftmost
and rightmost pseudo-zeros.

\subsection{Computing the leftmost functional pseudo-zero}
\vspace{-1mm}

Given an equation $E$ with variables $x_1,\ldots,x_n$.
The purpose of function $\ZA$ described in this section is to compute
the $i$-th functional leftmost pseudo-zero of $E$.
It is an adaptation of {\bf function} {\tt LeftNarrow} in \cite{vhlmyd97}.

Suppose function $\ZA$ is first called with arguments $a_0$ and
$b_0$ such that the $i$-th leftmost functional pseudo-zero, if it
exists, is contained in $[a_0,b_0]$.
It returns an interval $[a^\prime,b^\prime]$. If this interval is
empty, the algorithm has detected that no solution occurs in
$[a_0,b_0]$.
Otherwise, $[a^\prime,b^\prime]$
is the $i$-th functional leftmost pseudo-zero.

See Figure~\ref{ZERO1} for a definition of $\ZA$.  Here $f([a,b])$
denotes the result of evaluating $E$ with $B_j$ substituted for
variable $x_j$ for all $j \in \{1,\ldots,n\}\setminus\{i\}$ and
$[a,b]$ substituted for $x_i$.

\begin{figure}
\label{ZERO1}
\begin{tabbing}
xxxx \= xxxx \=       \kill
interval function $\ZA (a,b)$ \{         \\
     \> if $(0 \not \in f([a,b]))$ return $\emptyset$;  \\
     \> if ($[a,b]$ is canonical) return $[a,b]$;      \\
     \> //$[a,b]$ not canonical, so has midpoint      \\
     \> $ m := $ midpoint of $[a,b]$\/;                \\
     \> $I^\prime := \ZA(a,m);$                \\
     \> if ($I^\prime$ is empty) return $\ZA(m,b);$       \\
     \> return $I^\prime;$                            \\
\}
\end{tabbing}
\caption{
A definition of a function to compute a functional leftmost
pseudo-zero.
}
\end{figure}

\subsection{Computing canonical leftmost pseudo-solutions}
\vspace{-1mm}
The function $\ZB$ is  defined recursively  to compute the
relational leftmost  pseudo-zero; see Figure~\ref{ZERO2}.
\begin{figure}
\label{ZERO2}
\begin{tabbing}
xxxx \= xxxx \=       \kill
interval function $\ZB (a,b)$ \{                     \\
   \> if ($[a,b]$ is canonical) return $[a,b]$;      \\
   \> $ m := $ midpoint of $[a,b]$\/;                \\
   \> probe $\C$ with $x_i \leq m$;                  \\
   \> if (result of probing is failure) \{           \\
   \>   \>probe $\C$ with $m \leq x_i \wedge x_i \leq b$\/;  \\
   \>   \>if (result of probing is failure) return $\emptyset$; \\
   \>   \>return $\ZB(m,b)$\/;                       \\
   \> \}                                             \\
   \> $I^\prime := \ZB(a,m);$                        \\
   \> if ($I^\prime$ is empty) return $\ZB(m,b)$;    \\
   \> return $I^\prime;$                             \\ 
\}
\end{tabbing}
\caption{
A definition of a function to compute the $i$-th canonical leftmost
pseudo-solution. }
\end{figure}

It assumes and maintains the invariant that when $\ZB$ is called with
$a$ and $b$ as arguments, the interval $[a,b]$ is an $i$-th
leftmost
pseudo-solution. The result is $\emptyset$ when it has been proved
that no solution exists in $[a,b]$. If the result is not $\emptyset$,
then it is $[a_0,b_0]$ such that $[a_0,b_0] \subseteq [a,b]$ and $a_0
\leq b_0$, and $[a_0,b_0]$ is the $i$-th canonical leftmost
pseudo-solution.

\begin{definition}[Functionally truncated probing]
Let $\C$ be the constraint system associated with an equation system.
\emph{Functionally truncated probing} of $\C$ with a constraint $A$
refers to the result (failure or nonfailure) of an initial functional
segment of a trace of the constraint system $\C^\prime$ that results
from adding $A$ to $\C$.  The  result is nonfailure if $0$ is contained
in all intervals of variables occurring in the equality constraints in
$\C^\prime$; failure otherwise.
\end{definition}

If in $\ZB$ one would replace probing with functionally truncated
probing, then an algorithm would result that is for practical purposes
equivalent to $\ZA$.  That is, functional box consistency can be
computed with the same efficiency by constraint propagation on a system
of atomic constraints, provided that propagation is not chaotic, but
suitably structured and truncated.

\section{Conclusions}
\vspace{-1mm}

Following \cite{bmcv94,vhlmyd97} we have treated systems of
equations. Functional box consistency is easy to generalize to systems
containing both equalities and inequalities.  Relational box
consistency seems more general because it applies to all interval
constraint systems, not just to those that are derived from systems of
equalities and inequalities.

In \cite{bmcv94} {\tt Newton} was compared with CLP(Intervals) on the
Broyden Banded function.  CLP(Intervals) was observed to required time
exponential in the number of variables, whereas {\tt Newton} required
linear time.  This is indeed to be expected: CLP(Intervals) used for
pruning a single application of GCI.  Because of the weakness of such
pruning, the search tree reaches a significant depth. The size of such
a tree is exponential in the number of dimensions.

In \cite{bmcv94}, the observation was made that {\tt Newton} requires
no branching on this example. Hence no exponential behaviour is to be
expected.

It is now time to look beyond this particular example to those where
even with pruning as powerful as in {\tt Newton}, substantial
branching is necessary.
{\tt Newton} showed that more effort spent in pruning is rewarded by a 
reduction in branching in such a way that the total computation time
is much reduced. Thus there is a trade-off between time spent on
pruning and time spent on branching: at some point, additional effort
spent on pruning must stop being productive. This might suggest
replacing functional by relational box consistency.

There is a better method. Note that to compute both functional and
relational box consistency, one iterates all the way down to canonical 
intervals, the narrowest that the floating-point hardware allows. This 
is done to make the interval for one variable as narrow as
possible. Yet to compute box consistency one has to do this for all
variables repeatedly until no further narrowing is possible for any
variable. In the beginning, most the intervals for most of the
variables are still wide. While this is the case, it seems wasteful to
iterate in functions $\ZA$ or $\ZB$ all the way down to canonical
intervals: the convergence criterion should be adapted to the width of 
the other intervals.

In addition to this improvement, which applies both to $\ZA$ and to
$\ZB$, there is an improvement that applies to the latter alone.  As we
described probing here, propagation is completed to convergence. As
shown in Proposition~\ref{simul}, truncating the trace in propagation
to the initial functional segment, causes $\ZB$ to compute functional
box consistency. By truncating the trace less drastically, say, till
after the first cycle of a two-phase iteration, one obtains better
chance at getting failure in probing, yet avoids the negligible
reductions associated with the last phases of propagation to
convergence.  It seems worth investigating how many phases are optimal
in this respect.

\section{Related work}
\vspace{-1mm}
The routine {\tt absolve} in BNR Prolog uses probing to
find narrower intervals than a
single application of GCI can give.
The mechanism was discovered independently
by Chen and van Emden \cite{chnvnmdn97}, who called it
``hypernarrowing''.  It uses bisection to determine the greatest $m$
such that probing with $x \leq m$ gives failure.
In \cite{chnvnmdn97} ``hypernarrowing'' was used in optimization.  A
dramatic decrease in the number of function evaluations resulted.
\cite{chnvnmdn97} missed the connection between ``hypernarrowing'' and
box consistency.

Benhamou et al. \cite{bmcv94} noted the ineffectiveness of the
CLP(Intervals) {\tt solve} routine for nonlinear equations.
In response they
introduced box consistency and used it to achieve dramatic improvements
over the CLP(Intervals) {\tt solve}. In the version they introduced
(here called functional box consistency), they implicitly discard
constraint propagation, and only use interval arithmetic.

The {\tt HC4} algorithm of \cite{bggp99} is a propagation algorithm
where instead of individual contractions one applies an algorithm
called {\tt HCRrevise}, which similar to a two-phase iteration
truncated after the first cycle.


\end{document}